# In-plane anisotropy in biaxial ReS$_2$ crystals probed by nano-optical imaging of waveguide modes

Fabian Mooshammer[1,*], Sanghoon Chae[2,†], Shuai Zhang[1], Yinming Shao[1], Siyuan Qiu[1], Anjaly Rajendran[3], Aaron J. Sternbach[1], Daniel J. Rizzo[1], Xiaoyang Zhu[4], P. James Schuck[2], James C. Hone[2], and D. N. Basov[1,*]

[1]Department of Physics, Columbia University, New York, NY 10027, USA

[2]Department of Mechanical Engineering, Columbia University, New York, NY 10027, USA

[3]Department of Electrical Engineering, Columbia University, New York, NY 10027, USA

[4]Department of Chemistry, Columbia University, New York, NY 10027, USA

ABSTRACT: Near-field imaging has emerged as a reliable probe of the dielectric function of van der Waals crystals. In principle, analyzing the propagation patterns of subwavelength waveguide modes (WMs) allows for extraction of the full dielectric tensor. Yet previous studies have mostly been restricted to high-symmetry materials or narrowband probing. Here, we resolve in-plane anisotropic WMs in thin rhenium disulfide ($ReS_2$) crystals across a wide range of near-infrared frequencies. By tracing the evolution of these modes as a function of crystallographic direction, polarization of the electric field and sample thickness, we have determined the anisotropic dielectric tensor including the elusive out-of-plane response. The excitonic absorption at ~1.5 eV manifests itself as a clear backbending feature in the WM dispersion and a reduction of the quality factors as fully supported by numerical calculations. Our results extend the sensitivity of near-field microscopy towards biaxial anisotropy and provide key insights into the optoelectronic properties of $ReS_2$.

Van der Waals (vdW) crystals hold great promise for the field of nanophotonics[1,2]. Their layered nature induces a strong optical and electronic anisotropy between the in- and out-of-plane directions, thus giving rise to exceptional phenomena such as optical hyperbolicity[3] and the concomitant subwavelength confinement of light into waveguided polaritonic rays[4,5]. Further ramifications of the weak vdW coupling of the individual layers also raises the possibility of fabricating structures with arbitrary stacking angles[6,7]. In this manner, Coulomb correlations[8], lattice vibrations[9], and polaritons[10,11] have previously been tailored and novel phase transitions[12] have been designed.

To systematically exploit the strategies described above, a precise knowledge about the optoelectronic properties of vdW crystals is indispensable. Yet established spectroscopic ellipsometry of thin, exfoliated flakes remains challenging as the nanoscale sample thickness prevents a reliable measurement of the out-of-plane properties under normal incidence.

Meanwhile, optical responses have readily been accessed on the nanoscale with scattering-type scanning near-field optical microscopy[13–17] ($s$-SNOM) – even on layered materials[18]. Recently, in-plane anisotropy was resolved with antenna-assisted $s$-SNOM, albeit at the cost of patterning a metallic disk on top of the sample[19]. Conversely, waveguide modes (WMs) can propagate as cylindrical waves through vdW crystals without the need for special protocols for sample preparation beyond simple exfoliation. These electromagnetic waves confined to thin slabs comprise different types of modes with distinct polarizations and hence allow for selective access to either the in- or out-of-plane dielectric properties. As such, WMs are naturally sensitive to in-plane anisotropy. The expected deformation of the wavefronts from cylindrical into elliptical shape, can, in principle, be resolved by determining the characteristic wavelengths of the WMs along different high-symmetry directions of the vdW crystal.

Here, we investigate WMs in the prototypical in-plane anisotropic vdW crystal $ReS_2$ (refs. [20,21]). By imaging interference patterns of these electromagnetic modes close to the edges of thin (> 5 nm) exfoliated crystals (Figure 1a), we access the WM wavevector and hence the dielectric response along different crystallographic directions. A variation of the photon energy across the exciton resonance at ~1.5 eV reveals a pronounced backbending feature in the WM dispersion and a reduction of the WM quality factor induced by increased optical absorption. Our study expands the extraction of dielectric properties via WM nanoimaging to optically biaxial crystals, provides a direct measurement of the dielectric tensor of ReS$_2$ over a wide frequency range, and explores light-matter coupling in anisotropic vdW waveguides.

Rhenium-based transition metal dichalcogenides (TMDs) crystalize in a distorted 1T' structure[20,21] breaking the three-fold rotational symmetry known from the stoichiometrically related molybdenum or tungsten TMDs that are prevalent in the 2H phase. The top view of a monolayer of ReS$_2$ depicted in Figure 1b highlights the resulting in-plane anisotropy[20–23]. Particularly prominent are the Re-Re chains aligned along the *y*-direction. In bulk crystals, individual layers are stacked on top of each other but effectively remain decoupled[24]. As a result of this weak interlayer coupling, ReX$_2$ (X = S,Se) retains a direct band gap even in its bulk form[23,25] in contrast to the direct-indirect gap transition in MoX$_2$ and WX$_2$. Hence, rhenium-based TMDs are natural candidates for optoelectronic applications such as energy harvesting. Furthermore, this material class holds great potential for photocatalysis and sensor applications[20,21]. Especially at low temperatures, the optical absorption of ReX$_2$ is highly anisotropic[25–28], which represents the basis for polarization-resolved photodetectors, for example. While the optical properties of ReS$_2$ have previously been studied with photoluminescence[24,26], reflection contrast[29] or absorption[30,31] measurements, its dielectric tensor and especially its out-of-plane response have mostly remained elusive.

RESULTS

**Nanoimaging of waveguide modes.** The concept of WM nanoimaging is illustrated in Figure 1a. Near-infrared continuous-wave radiation is focused onto the metallized tip of an atomic force microscope. The light is coupled into evanescent modes at the sharp apex of the near-field probe, which can efficiently excite WMs. Subsequently, the wavefronts of these modes propagate away from the tip in a cylindrical fashion until they encounter the edges of the sample. There, the WMs are predominantly coupled out – in stark contrast to polaritons that are typically reflected in such a scenario due to their much stronger field confinement and the associated increased momentum mismatch with vacuum[32]. At the same time, a fraction of the incident radiation is also elastically scattered from the tip as in other *s*-SNOM experiments. The light from both sources – tip and sample edge (blue arrows in Figure 1a) – is collected, superimposed and detected in a pseudo-heterodyne scheme (see Methods). For background-free images, the signals are additionally demodulated at higher harmonics $n \geq 4$ of the tip tapping frequency to isolate the amplitude $s_n$ and phase $\varphi_n$ (which we do not consider here) of the scattered light.

By virtue of the interference between the light emanating from the tip and the sample edges, fringe patterns in the scattered amplitude $s_n$ emerge along the perimeter of the microcrystal. For a representative sample with a thickness $d = 156$ nm (Figure 1c) this phenomenon is illustrated by the map of $s_5$ obtained with an incident wavelength of $\lambda = 950$ nm (photon energy: 1.31 eV), as shown in Figure 1d. Clearly, different fringe periodicities are observed along orthogonal in-plane directions. Yet such images need to be interpreted with caution as the interference pattern caused by light emitted from the tip and the sample edge is also dictated by the angle of these sources relative to the direction of the detector. As a result, different fringe periodicities may be observed for different sample orientations with respect to the incident light[32–34] even along the same edge of the microcrystal. For the same reason, the interference patterns next to the left-hand and right-hand vertical edges drastically differ in wavelength even though they are generated by the same WMs. Generally, the interference is governed by the angles γ and δ as defined in Figure 1a, which represent the out-of-plane angle of the incident light and the relative in-plane orientation of the sample edge, respectively. In ref. [34] an analytical expression has been derived based on these considerations, which allows for a faithful extraction of the wavevector $k$ of the WMs from real-space images. In Supporting Information (SI) Section 1, we outline the key equations and verify this formalism by analyzing fringe patterns obtained for various rotations of the ReS$_2$ crystal. Once the correct wavevector $k$ has been identified for a given photon energy with Equation S1, the dielectric properties can be inferred as previously reported for various molybdenum[32–38] and tungsten-based[34,39] dichalcogenides or inorganic perovskites[40]. This approach has previously been demonstrated to even resolve transient changes to the optical properties on femtosecond timescales resolve pulsed laser sources[34,39].

Due to the strong in-plane anisotropy, ReS$_2$ preferentially cleaves along specific crystallographic directions upon mechanical exfoliation[22,23]. Most of our samples (see Figure S2) are stripe-like with two parallel edges that are several micrometers long and run parallel to Re-Re chains in the *y*-direction (Figure 1b). To corroborate this, we performed Raman spectroscopy[22,23,41] with linearly polarized light for various in-plane polarizations. A representative Raman spectrum is shown in Figure 1e. Here, mode V at ~211 cm$^{-1}$ is particularly sensitive to the crystallographic orientation[22] of ReS$_2$. The polar plot of the peak intensity of mode V in Figure 1f reveals clear maxima when the linear polarization of the excitation laser is aligned with the vertical, parallel edges of the flake (compare Figure 1c). Thus, the Raman data confirm that these sample boundaries are indeed aligned with Re-Re chains.

**Waveguide mode dispersion.** Next, we analyze the waveguide modes propagating along the *x*-direction in more detail. Figure 2a shows linecuts of $s_5$ extracted from scans similar to Figure 1d for a range of photon energies from 1.25 eV to 1.72 eV. In each case, numerous interference fringes are observed close to the sample edges. Strikingly, the wavelength and the propagation length of the WMs is significantly reduced with increasing photon energy.

To quantify these phenomena, we perform a Fourier analysis of the linecuts of $s_5$. The Fourier traces (Figure 2b) account for the geometrical dependence[34] of the fringe periodicity on the actual wavevector $k_x$ of the WMs. Additionally, components generated by the step-like increase of $s_5$ at the sample edge are suppressed by a window function[33] (SI Section 1). We identify multiple signatures in these traces, where $k_x$ is given in units of the free-space wavevector $k_0$ of the incident light: i) The dispersionless air mode at $k_x \sim k_0$. ii) the fundamental WM at $k_x \sim 2k_0$ marked by the triangles, and iii) two additional higher-order modes labeled by the squares and circles, respectively, which are most pronounced for photon energies <1.5 eV.

To identify the nature of the three WMs, we calculate the mode dispersion (Figure 2c) hosted by the air/ReS$_2$/SiO$_2$/Si structure via the imaginary part of the Fresnel reflection coefficients $\text{Im}(r_s + r_p)$ obtained with the transfer-matrix formalism[42]. The three branches in the dispersion correspond to the zero-order transverse magnetic (TM$_0$), as well as the first- and zero-order transverse electric (TE$_1$ and TE$_0$) modes. The field profiles of these modes are depicted in Figure 2d. Importantly, the electric fields feature components of drastically different polarizations. Whereas TM modes contain electric fields along the out-of-plane direction ($E_z$) and along the direction of propagation ($E_x$), the TE modes only feature the orthogonal in-plane component ($E_y$). Consequently, TM and TE modes can efficiently be excited by p-polarized or s-polarized light[43], respectively. Since the electric fields at the tip apex are dominated by out-of-plane components for the experiments with p-polarized light in Figure 2a,b, the TM$_0$ modes dominate the interference pattern and the amplitude spectra in these cases. By extracting the wavevectors of the modes for each photon energy, we experimentally trace the full dispersion from the near infrared to the boundary of the visible range and find very good agreement with $\text{Im}(r_s + r_p)$. Here, the dielectric response of ReS$_2$ along the in-plane directions ($\varepsilon_x$, $\varepsilon_y$) has been modelled by Lorentzian oscillators, whereas a constant value was assumed for the out-of-plane axis ($\varepsilon_z$) due to the effective decoupling of the layers (compare Figure 3b). The resonance at ~1.5 eV can be identified as the absorption by the exciton, whose binding energy amounts to tens of meV, rendering the quasiparticle stable at room temperature[27,44]. In the WM dispersion, the exciton resonance gives rise to increased levels of damping[32,34] as well as a pronounced backbending feature that puts our samples on the verge of strong light-matter coupling[38] with a Rabi splitting on the order of 0.1 eV (Figure S3). Experimentally, these signatures manifest themselves as a dispersive behavior of $k_x$ and a reduction of the amplitudes of the modes in Figure 2b close to the exciton resonance – a feature that is more pronounced for the TE modes as discussed later on.

**Quantifying the dielectric response of ReS₂.** Analogous experiments along the *y*-direction, for which WMs are coupled out by the horizontal edge of the flake in Figure 1c,d, yield qualitatively similar dispersions (SI Sections 1 & 3). For this configuration the wavevectors of the TE modes are noticeably shifted towards smaller values (Figure 3a). This behavior is in line with the expected larger refractive index along the rhenium chains[29,45] (*y*-direction) given that the TE modes feature electric fields orthogonal to their direction of propagation (schematics in Figure 3a). In contrast, the TM modes are only weakly susceptible to the in-plane anisotropy of ReS₂ owing to their additional out-of-plane electric field component (Figure S1 & S3).

The distinct polarizations of different types of WMs can be exploited to extract the dielectric response $\epsilon = \epsilon' + i\epsilon''$ from interference fringes with wavevectors along different crystallographic directions. To this end, we consider the transcendental equations for the TE and TM waveguide modes introduced in ref. [33], and generalize them to arbitrary in-plane directions (solid lines in Figure 3a) using the index ellipsoid formalism (see Methods).

By numerically minimizing the discrepancy between experimental WM dispersions (as in Figure 2c) and the predictions of Equations 2-4 (see Methods), we retrieve the principal components of the dielectric tensor (see Figure 3b). Due to the decoupled nature of the ReS₂ layers, the out-of-plane response $\varepsilon_z$ essentially takes a constant value in our spectral range. Conversely, we indeed extract Lorentzian resonances along the in-plane directions of ReS₂. These findings are analogous to bulk MoS₂, whose out-of-plane refractive index exhibits negligible dispersion around the main exciton resonances, while the in-plane components are highly dispersive[37]. Generally, the transition dipole moments are predominantly oriented along the in-plane direction for bright excitons in vdW semiconductors[46]. Note that our retrieval procedure (Figure S4) does not require prior assumptions about the spectral shape of any principal component of the dielectric tensor. Along the *y*-direction, the static dielectric constant ($\epsilon^0$) is larger by ~24 %, giving rise to the ellipsoidal shape of the polar plot of the WM wavevectors in Figure 3a.

The imaginary parts of the dielectric responses can be assessed by evaluating the losses of the WMs. We extract the propagation lengths $L$ of the $TM_0$ WMs by fitting line profiles of $s_n$ with an exponentially damped, oscillating function (Figure S5). For increasing photon energy, $L$ decreases (black symbols) in unison with the wavelength of the incident light. This behavior is even present for dispersionless dielectric responses, which renders the impact of the exciton resonance more subtle. Consequently, we instead consider the quality factors $Q$ (red symbols). Thus, we reveal enhanced losses of the WMs close to the exciton resonance around ~1.5 eV. Due to the dominant out-of-plane polarization of the $TM_0$ modes and the in-plane dipole of the exciton, the losses stay moderate even for energies close to the resonance in contrast to isotropic media such as perovskites[40] or dye films[47]. Along these lines, the TE modes experience stronger damping due to the exciton in $ReS_2$ compared to the TM counterparts, which renders the observation of the former beyond >1.5 eV more challenging. An analytical model based on approximating the WMs as cylindrical waves (SI Section 5), which takes the dielectric response in Figure 3b as input, can adequately reproduce the experimental observations. The slight overestimation of $Q$ for photon energies above ~1.6 eV is attributed to the onset of the absorption of $ReS_2$ beyond the electronic gap[30,31], which is not included in the Lorentzian in-plane dielectric response.

Whereas the real part of the dielectric response ($\epsilon'$) is directly linked to the periodicity of the interference pattern and can therefore be extracted very accurately, the determination of the imaginary part ($\epsilon''$) governing the decay profile of the WMs is a much more subtle task (Figure S5). Within the resulting error margins, the losses and quality factors of the $TM_0$ mode along both $x$- and $y$-directions of the microcrystals (compare Figure 3c) do not exhibit clear anisotropy. This observation is, however, in line with the comparable oscillator strengths of the resonances in $\epsilon_x$ and $\epsilon_y$ at ambient conditions and the negligible losses below the exciton absorption.

**Probing in-plane anisotropy.** Finally, we use a complementary approach to resolve the in-plane anisotropy of ReS$_2$ with even higher accuracy. To this end, we exclusively probe TE$_0$ modes in thin flakes (5 nm < $d$ < 60 nm). To excite these modes more efficiently[43] than in Figure 2, we switch to s-polarized light at a wavelength of λ = 950 nm. The upper panels of Figure 4a,b show representative maps of $s_4$ for $d$ = 30 nm and $d$ = 59 nm. Clear interference fringes are observed as far as several microns away from the sample edge as highlighted in the line traces in the lower panels. As $d$ increases, the wavevector $k_x$ of the WM is shifted from ~1.9 $k_0$ to ~2.9 $k_0$. Tracing the evolution of $k_x$ across a wide range of sample thicknesses $d$ therefore facilitates a precise determination of the optical response. Figure 4c summarizes the results obtained analogously to Figure 4a,b and also includes the wavevectors of TE$_1$ and TM$_{0/1}$ modes recorded on thicker samples. Repeating the nanoimaging experiments along different edges of the thin microcrystals, we can even map out the full thickness dependence of the TE$_0$-mode dispersion along all in-plane directions (Figure 4d). Using the framework of Equations 2-4 described in the Methods section, we can reproduce the entire data depicted in Figure 4 with only three parameters, that is, ($\epsilon'_x$, $\epsilon'_y$, $\epsilon'_z$) = (18.0, 21.3, 6.0). These values are consistent with the analysis in Figure 3b (see black symbols). As discussed above, the imaginary parts of the dielectric function $\epsilon''$ are more challenging to quantify. Based on the findings in Figure 3b, we expect a transparency window of the in-plane response up to the exciton resonance and a negligible out-of-plane absorption across the investigated near-infrared spectral range similar to MoS$_2$ crystals[37].

DISCUSSION AND CONCLUSION

The excellent agreement between the numerical modelling and the experimental data covering an extended parameter space demonstrates the capabilities of near-field microscopy for characterizing anisotropic optical responses through WMs. The validity of our approach is further corroborated by a comparison to the literature: $\epsilon'_x$ almost perfectly matches the result of an ellipsometry study[44] (black dashed line in Figure 3b), and the retrieved $\epsilon'_z$ lies exactly in the range between the static ($\epsilon^0_z$) and optical ($\epsilon^\infty_z$) dielectric response predicted by density functional theory[48]. Moreover, a complementary comparison of the near-field results to polarization-resolved far-field reflection experiments (see Methods and Figure S6) – similar to the correlative studies in ref. [49] – corroborates the accuracy of the retrieved in-plane dielectric response. In particular, the key characteristics of the exciton resonance and the difference in dielectric background between *x*- and *y*-directions observed in far-field reflection are consistent with the data in Figure 3b.

This allows us to reliably quantify the birefringence of ReS$_2$ between in- and out-of-plane directions via the difference in refractive index $\Delta n$. Due to the vdW stacking of the individual layers, a giant optical anisotropy ($\Delta n > 1.8$, compare Figure 3b) emerges in the near infrared similar to molybdenum and tungsten dichalcogenides[37]. This feature renders ReS$_2$ promising for photonic applications such as waveguides with extreme skin depth[37] or low-loss Mie resonators[50], where the in-plane anisotropy can serve as an additional tuning knob for tailoring electromagnetic fields on subwavelength scales.

In conclusion, we have imaged interference patterns created by waveguide modes (WMs) in ReS$_2$ flakes from the near infrared to the boundary of the visible spectral range. Mapping the evolution of the WM wavevector as a function of photon energy, sample thickness, or polarization of the incident electric field, has allowed for a retrieval of the anisotropic dielectric response of ReS$_2$ along all crystallographic directions. The exciton resonance at ~1.5 eV induces a backbending of the WM dispersion and a reduction of the propagation lengths of the waveguide modes – a measure for the losses of the material.

In the future, the potential directional-dependent modal birefringence[51] between TE$_1$ and TM$_0$ modes could be explored in rhenium-based dichalcogenides. In general, our approach is applicable to a large zoo of anisotropic layered materials with low symmetry[52], whose exceptional properties have yet to be fully unveiled. In combination with cryogenic near-field microscopy[53], exciton polaritons[54] that are anisotropic and self-hybridized[55] or highly confined[56] could thus be imaged. Finally, our technique is also easily transferrable to the visible[40] or mid-infrared range[57] paving the way for a full characterization of novel materials for nanophotonic applications across various spectral ranges.

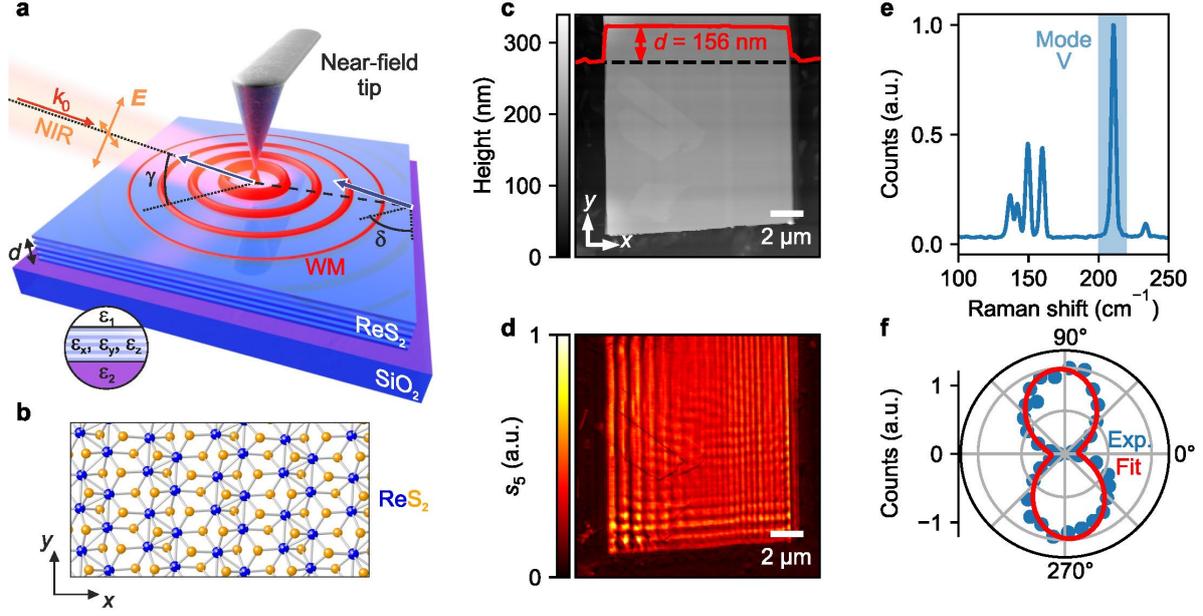

**Figure 1.** Infrared nanoscopy and nanoimaging of waveguide modes in $ReS_2$ crystals. (a) Schematic of the experimental setup. Near-infrared (NIR) radiation with wavevector $k_0$ and vertically (p-) or horizontally (s-) polarized electric field $E$ (orange arrows) is focused onto a metallic probe under an angle of $\gamma = 30°$ with respect to the sample surface. Via the evanescent fields at the sharp tip apex (see Methods for an alternative mechanism), the radiation is coupled into waveguide modes (WMs) inside a slab of the $ReS_2$ crystal. The detected fields are a superposition of the light scattered into the far field by the tip and by the sample boundaries (blue arrows) and hence depend on the angle $\delta$ enclosed by the projection of $k_0$ and the sample edge. The $ReS_2$ microcrystals with thickness $d$ have been mechanically exfoliated onto a $SiO_2$/Si substrate. Inset: layered dielectric response of the structure encompassing air ($\varepsilon_1$), $ReS_2$ ($\varepsilon_x$, $\varepsilon_y$, $\varepsilon_z$), and $SiO_2$ ($\varepsilon_2$). (b) Top view of the ball-and-stick model of a $ReS_2$ monolayer for which the characteristic Re-Re chains are oriented along the $y$-direction. (c) Topography of a representative sample with a thickness of $d\sim156$ nm recorded with atomic force microscopy. The red line profile was extracted along the black dashed line. The parallel, nominally vertical edges of the flake indicate the $y$-direction of the crystal. (d) Map of the near-field amplitude $s_5$ of the region shown in (c) for an incident wavelength of $\lambda = 950$ nm (photon energy: 1.31 eV) recorded with $\delta = 118°$. (e) Representative Raman spectrum obtained on the flake in (c,d). The blue-shaded region highlights mode V (compare ref. [22]) at an energy of ~211 cm$^{-1}$. (f) Intensity of mode V as function of the orientation of the linearly polarized excitation with respect to the crystallographic axes. The maxima of the fit (red line) with Equation 1 to the experimental data (symbols) indicates the $y$-direction (Re-Re chains) of the $ReS_2$ (see Methods).

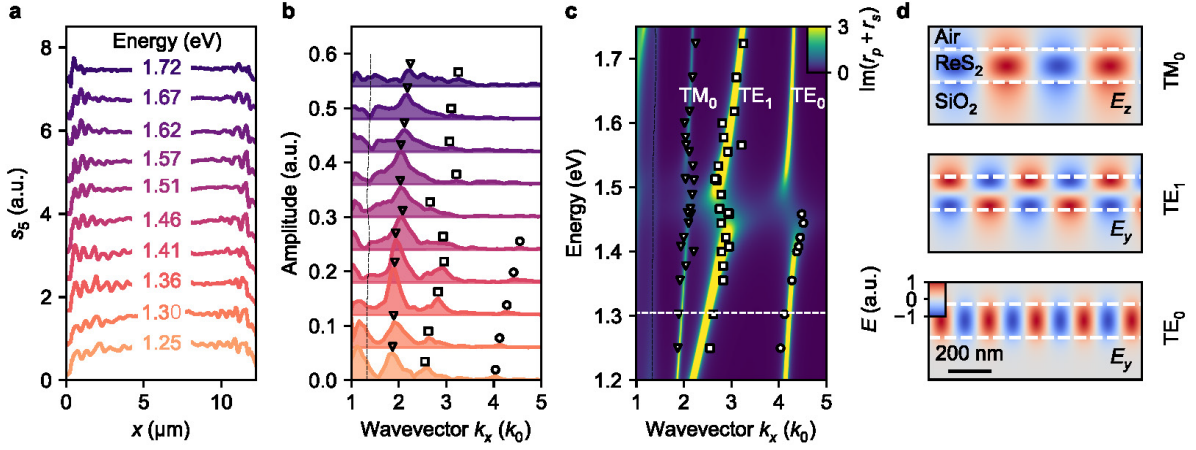

**Figure 2.** Measuring the dispersion of the waveguide modes. (a) Line profiles of the near-field amplitude $s_5$ for a series of photon energies. The data were obtained with δ = 151°, extracted along the x-direction in Figure 1c. (b) Fourier analysis of the line profiles in (a) using the same color code. For clarity, the data in (a,b) are vertically offset by 0.75 and 0.06, respectively, for each trace. A window function was applied to suppress spectral components generated by the variation of $s_5$ at the sample edges before performing a geometric correction of the wavevector. For details see refs. [33,34] and SI Section 1. The peaks in the spectra (symbols) represent different types of WMs. The wavevector $k_x$ is given in units of the free-space wavevector $k_0$. The vertical dashed line traces the modes inside the $SiO_2$ light cone extracted from panel (c). Note that the amplitudes for $k_x$ much smaller than the $TM_0$ mode are affected by the windowing process (see SI Section 1). (c) Map of wavevector $k_x$ versus photon energy derived from the data in (a,b) and from additional datasets recorded on the flake in Figure 1. The same symbols as in (b) were used. The sum of the imaginary parts of the Fresnel reflection coefficients $\text{Im}(r_p + r_s)$ indicates the dispersion of the transverse magnetic ($TM_0$) and transverse electric ($TE_1$ and $TE_0$) modes based on the fitted dielectric response depicted in Figure 3b. The white, dashed line highlights the photon energy used for the field distributions in (d). For better visibility of the dispersion, an imaginary part of 0.1i was added to all dielectric responses. (d) Electric field distributions of the air/$ReS_2$/$SiO_2$/Si structure for the $TM_0$, $TE_1$, and $TE_0$ modes along the x-direction. The TM and TE modes feature orthogonal electric field components $E_{x/z}$ and $E_y$, respectively. The numerical simulations in (c,d) were performed by adapting the code provided in ref. [42].

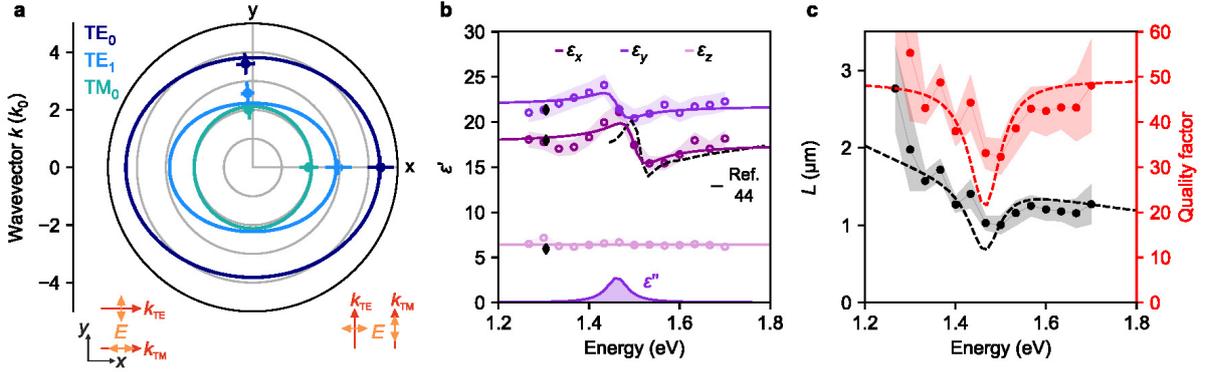

**Figure 3.** Anisotropic dielectric response and quality factor. (a) Polar plot of the wavevectors $k$ of the WMs in the ReS$_2$ microcrystal depicted in Figure 1c,d along different in-plane directions at a photon energy of 1.4 eV. The solid lines represent the numerical solutions of Equations 2-4 (see Methods) using the dielectric response in (b). The radial error bars were determined by fitting the peaks in the Fourier analyses with Gaussian functions and extracting their typical widths. The schematic arrows indicate representative in-plane polarizations of the electric field $E$ for TE and TM WMs with wavevectors $k_{TE}$ and $k_{TM}$, respectively. (b) Tensor elements $\varepsilon_x$, $\varepsilon_y$, $\varepsilon_z$ of the real part $\varepsilon'$ of the dielectric function of ReS$_2$ determined by WM imaging. The symbols represent experimental data extracted from the mode dispersions along the crystallographic $x$- and $y$-directions of various samples. The shaded intervals indicate the relative errors determined by evaluating the discrepancy between the experimental and the retrieved wavevectors (see Figure S4). Solid lines: fits with Lorentzians/constant value for the in-plane/out-of-plane components. Shaded region: imaginary part $\varepsilon''_y$ of the fit to the dielectric response. Dashed line: ellipsometry data from ref. [44]. Black symbols: dielectric response determined with the thickness dependence in Figure 4c,d. (c) Propagation length $L$ (black) and quality factor $Q$ (red) of the TM$_0$ WMs along the $y$-direction (see Figure S3) determined by fitting the line profiles (see Figure S5) with a function accounting for the geometrical decay and the damping induced by the imaginary part of the wavevector $k_2 = (2L)^{-1}$. The ratio of the real part of the corrected wavevector $k_1^{corr}$ and $k_2$ was used for calculating $Q = k_1^{corr}/k_2$. The error margins were determined through the fitting procedures. The black and red line shows the results of the theoretical model (compare Equations S5 & S6) using the dielectric function in (b) as input and assuming an additional imaginary part of the dielectric function along the out-of-plane direction of 0.26 (see SI Section 5).

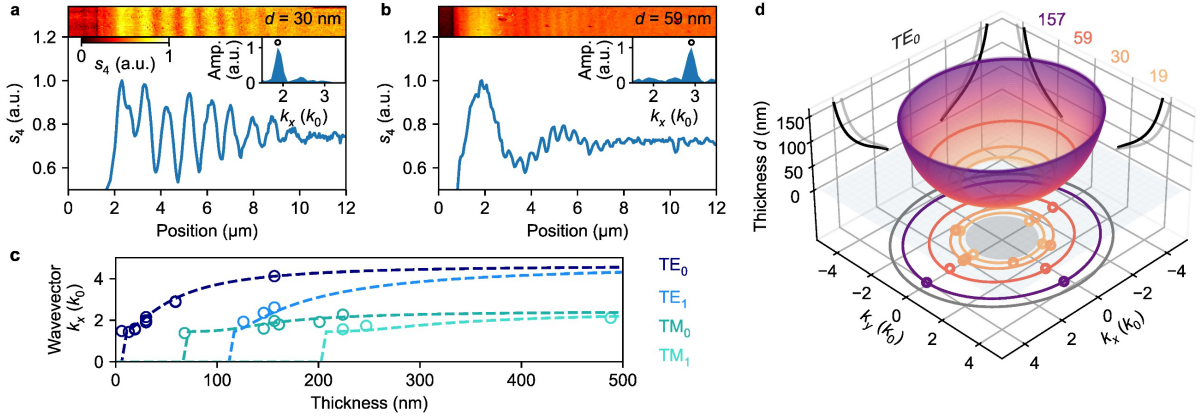

**Figure 4.** Thickness dependence and anisotropic TE$_0$ mode dispersion. (a,b) Upper panels: Near-field amplitude $s_4$ obtained with s-polarized s-SNOM on ReS$_2$ flakes with thicknesses $d = 30$ nm (a) and $d = 59$ nm (b), respectively. Lower panels: Line trace of $s_4$ determined by averaging the scans along the vertical direction ($\delta = 75°$). Insets: Amplitudes of the spectral components of the line traces (extracted in analogy to Figure 2b) in units of the free-space wavevector $k_0$. (c) Wavevectors $k_x$ for ReS$_2$ flakes with different thicknesses. The dashed lines indicate the theoretically predicted wavevectors for the TM$_{0/1}$ and TE$_{0/1}$ modes. The numerical calculations in (c,d) are based on Equations 2-4 (see Methods) with the dielectric function indicated by the black symbols in Figure 3b. (d) Theoretical, anisotropic dispersion of the TE$_0$ mode for flakes with different thicknesses $d$ depicted by the colored surface. Projections onto the x- and y-directions are shown in black on the side panels. On the $k_x$ ($k_y$) plane, the light gray lines represent the dispersion along $k_y$ ($k_x$) highlighting the in-plane anisotropy. For four representative thicknesses $d$, the respective colored contours are also projected onto the $k_x$-$k_y$-plane for a comparison with the experimental data (symbols) obtained for various crystallographic directions set by the edges of the flakes. Due to the cleavage tendency[23] of ReS$_2$ upon exfoliation, mostly discrete steps of 60° or 120° are accessible (Figure S2). The grey disk and circle indicate the cutoff by the refractive index of the substrate (small $d$) and the ReS$_2$ (infinite $d$), respectively.

METHODS

**Sample preparation and characterization.** Thin flakes of $ReS_2$ were exfoliated from a bulk crystal (www.hqgraphene.com) onto $SiO_2$/Si using the standard scotch tape method. Since the samples were obtained through multiple exfoliation processes on different substrates, the flakes are oriented randomly with respect to one another, resulting in different angles δ in consecutive experiments (compare Figure 1 and Figure 4). Using an optical microscope, flakes with parallel edges several micrometers in length were identified. To corroborate that these edges coincide with the Re-Re chains (*y*-direction; as defined in Figure 1b), Raman spectroscopy was performed on representative samples (compare Figure 1e,f). To this end, a linearly polarized cw-laser at a wavelength of 532 nm in a commercial Raman microscope (Bruker Senterra II) was used. The frequency axis was calibrated using the silicon phonon at 520 cm$^{-1}$. For the angle-resolved measurements in Figure 1f, the sample was placed on a rotation stage and the angle $\varphi$ was adjusted in steps of 10°. The scattered light was detected without analyzing the polarization and hence the total intensity $I(\varphi)$ of mode V (ref. [22]) was extracted for each sample orientation. According to ref. [58], the contributions of the light with parallel and perpendicular polarization with respect to the excitation, lead to a total Raman intensity:

$$I(\varphi) \propto u^2 \cos^2(\varphi - \varphi_0) + w^2 \sin^2(\varphi - \varphi_0) \\ + 2v(u+w)\sin(\varphi - \varphi_0)\cos(\varphi - \varphi_0). \tag{1}$$

In our case, *u*, *v*, and *w* serve as fitting parameters determining the functional dependence of $I(\varphi)$, whereas $\varphi_0$ alone sets the relative orientation of the crystallographic high-symmetry axes.

**Near-field experiments.** The nanoimaging of the waveguide modes was performed with a commercial scattering-type scanning near-field optical microscope (*s*-SNOM, neaSNOM by Neaspec GmbH). The widely tunable output (wavelength range: ~700 – 1000 nm) of a continuous-wave Ti:sapphire laser (SolsTiS by M Squared Lasers Inc.) is focused onto the sharp metallic probe of an atomic force microscope using a parabolic mirror. The near-field microscope is operated in tapping mode with typical tapping amplitudes of ~50 nm of the cantilever at driving frequencies of ~75 kHz. The radiation backscattered from the tip-sample system is collected by the same parabolic mirror and superimposed with a reference beam for pseudo-heterodyne interferometric detection[59]. For background-free imaging, the signal amplitudes $s_n$ of the scattered radiation are additionally demodulated at harmonics $n$ of the tapping frequency. We found that in contrast to the terahertz[60–62] and mid-infrared ranges[15,16], $n \geq 4$ is required for the visible and near-infrared spectral range in most experiments. For s-polarized light, the signal-to-noise ratio is decreased due to a weaker coupling to the evanescent fields at the tip apex. As a result, we sometimes resorted to $n = 3$ in such cases, while ensuring that the fringe pattern was the same as for higher harmonics (compare ref. [40]). For further details regarding the experimental setup, see ref. [63].

**Transcendental equations and index ellipsoid formalism.** The transcendental equations for the TE and TM waveguide modes along the *x*-direction with wavevector $k_x$ were introduced in ref. [33]:

$$\sqrt{\epsilon_y \times k_0^2 - k_x^2}\, d = \tan^{-1}\left(\frac{\alpha_1}{\sqrt{\epsilon_y \times k_0^2 - k_x^2}}\right) + \tan^{-1}\left(\frac{\alpha_2}{\sqrt{\epsilon_y \times k_0^2 - k_x^2}}\right) + n\pi, \quad (2)$$

and

$$\sqrt{\frac{\epsilon_x}{\epsilon_z}(\epsilon_z \times k_0^2 - k_x^2)}\, d = \tan^{-1}\left(\frac{\epsilon_x \alpha_1}{\epsilon_1 \sqrt{\frac{\epsilon_x}{\epsilon_z}(\epsilon_z \times k_0^2 - k_x^2)}}\right) + \tan^{-1}\left(\frac{\epsilon_x \alpha_2}{\epsilon_2 \sqrt{\frac{\epsilon_x}{\epsilon_z}(\epsilon_z \times k_0^2 - k_x^2)}}\right) + m\pi. \quad (3)$$

Here, $\alpha_i = \sqrt{k_x^2 - k_0^2 \times \epsilon_i}$ (i = 1,2) where $\epsilon_1 = 1$ and $\epsilon_2 = 2.10$ (ref. [37]) correspond to the dielectric response of the superstrate air and substrate SiO$_2$, respectively. In contrast, $\epsilon_x$, $\epsilon_y$, and $\epsilon_z$ characterize the dielectric response of ReS$_2$ (inset in Figure 1a). As in the majority of previous works[28,45], we do not consider off-diagonal elements of the dielectric tensor, which arise due to the trigonal crystal structure of ReS$_2$ but are negligible in magnitude[48]. The integers n and m are the order numbers of the TE$_n$ (Equation 2) and TM$_m$ (Equation 3) modes, respectively, and represent the number of nodes along the out-of-plane direction in the distributions of the electric field (Figure 2d). In essence, the dispersions resulting from Equations 2 & 3 trace the maxima in maps of $\text{Im}(r)$ while reducing the numerical effort.

This formalism can be generalized to arbitrary in-plane directions (solid lines in Figure 3a) characterized by an angle θ with respect to the *x*-axis. The index ellipsoid formalism yields the following equation for the dielectric response $\epsilon_{IP}(\theta)$:

$$\frac{1}{\epsilon_{IP}(\theta)} = \frac{\cos^2\theta}{\epsilon_x} + \frac{\sin^2\theta}{\epsilon_y} \qquad (4)$$

Due to the orthogonal polarizations of the electric fields of TM and TE modes, $\epsilon_x \to \epsilon_{IP}(\theta)$ and $\epsilon_y \to \epsilon_{IP}(\theta + \pi/2)$ in Equations 2 & 3 in this case.

**Alternative origin of the interference fringes.** In an alternative scenario compared to the one discussed in the main text, the interference patterns in the scattered amplitude $s_n$ could also originate from incident light that is directly coupled into WMs at the sample edges. After a propagation of the modes towards the tip, and an outcoupling to the far field, interference with the light directly scattered from the tip would generate fringe patterns in $s_n$. Yet both mechanisms ("edge outcoupling" and "edge incoupling") are expected to yield the same periodicity and hence the correct wavevector of the WMs according to our theoretical model described in SI Section 1. An investigation of the visibility of the interference fringes – a direct measure for the coupling efficiency of light from the far field into WMs (and vice versa) at the tip and at the sample edge – as a function of the sample orientation should facilitate a clear distinction of the two cases in the future. A recent report (ref. [37]) suggests, however, that the "edge outcoupling" scenario is the dominating one.

**Broadband reflection measurements.** The reflectance spectra discussed in SI Section 6 were obtained using a Hyperion 2000 microscope coupled with a Bruker FTIR spectrometer (Vertex 80V), equipped with a tungsten halogen lamp and a silicon detector. The polarized light was focused on the sample using a x36 objective using an aperture size smaller than the sample dimensions.

ASSOCIATED CONTENT

**Supporting Information**

The Supporting Information is available free of charge at

Extraction of wavevectors and geometric corrections; representative topography of additional flakes; backbending of the dispersion along the y-direction; retrieval of the dielectric function; propagation lengths and quality factors; comparison to far-field reflection (PDF)

**Data availability**

The data sets generated during and/or analyzed during the current study are available from the corresponding author upon reasonable request.

AUTHOR INFORMATION

**Corresponding Authors**

Fabian Mooshammer − Department of Physics, Columbia University, New York, New York 10027, United States; Email: fhm2115@columbia.edu

Dimitri N. Basov − Department of Physics, Columbia University, New York, New York 10027, United States; Email: db3056@columbia.edu

**Present Address**

[†]Sanghoon Chae − School of Electrical and Electronic Engineering, School of Materials Science and Engineering, Nanyang Technological University, Singapore 639798, Singapore.

**Author Contributions**

S.C., A.R. and J.C.H. fabricated the samples. F.M. performed the near-field experiments with support by S.Z., Y.S., A.S. and D.J.R. F.M. analyzed the data, characterized the samples with Raman spectroscopy, and performed the numerical simulations. Y.S. and S.Q. recorded the reflection spectra. F.M. and D.N.B. wrote the initial version of the manuscript. D.N.B. is the principal investigator and helped direct the course of the research. All authors contributed to the discussions and have reviewed and given approval to the final version of the manuscript.

**Notes**

The authors declare no competing interest.

**Funding Sources**

Research at Columbia is supported as part of Programmable Quantum Materials, an Energy Frontier Research Center funded by the U.S. Department of Energy (DOE), Office of Science, Basic Energy Sciences (BES), under award no. DE-SC0019443. D.N.B is the Gordon and Betty Moore Foundation's EPiQS Initiative Investigator no. 9455. F.M. gratefully acknowledges support by the Alexander von Humboldt Foundation.